\documentclass[preprint,aps,tightenlines,showpacs,nofootinbib]{revtex4}
\usepackage{latexsym}
\usepackage[dvips]{graphicx}
\newcommand{\beq}{\begin{eqnarray}}
\newcommand{\eeq}{\end{eqnarray}}
\newcommand{\eq}{eqnarray}

\newcommand{\al}{{\alpha}}
\newcommand{\be}{{\beta}}
\newcommand{\ci}{\cite}
\newcommand{\ga}{{\gamma}}

\newcommand{\ep}{{\epsilon}}

\newcommand{\de}{{\delta}}

\newcommand{\la}{{\lambda}}
\newcommand{\La}{{\Lambda}}

\newcommand{\om}{{\omega}}

\newcommand{\pa}{{\partial}}
\newcommand{\no}{{\nonumber}}
\newcommand{\f}{\frac}
\newcommand{\ra}{\rightarrow}

\begin{document}

\preprint{arXiv:0910.1917v4 [hep-th]}

\title{Remarks on the Scalar Graviton Decoupling and Consistency of Ho\v{r}ava Gravity}

\author{Mu-In Park\footnote{E-mail address: muinpark@gmail.com}}

\affiliation{ Research Institute of Physics and Chemistry, Chonbuk
National University, Chonju 561-756, Korea }

\begin{abstract}
Recently Ho\v{r}ava proposed a renormalizable gravity theory with
higher derivatives by abandoning the Lorenz invariance in UV. But
there have been confusions regarding the extra scalar graviton mode
and the consistency of the Ho\v{r}ava model. I reconsider these
problems and show that, in the Minkowski vacuum background, the
scalar graviton mode can be consistency decoupled from the usual
tensor graviton modes, by imposing the (local) Hamiltonian as well
as the momentum constraints.

\end{abstract}


\maketitle

\newpage

%\section{Introduction}
Recently Ho\v{r}ava proposed a renormalizable gravity theory with
higher spatial derivatives (up to sixth order) in four dimensions
which reduces to Einstein gravity with a {\it non-vanishing}
cosmological constant in IR but with improved UV behaviors by
abandoning the Lorentz invariance from non-equal-footing treatment
of space and time \ci{Hora:08,Hora}.  Due to lack of full
diffeomorphism, some extra graviton modes are expected generally but
there have been confusions regarding the extra modes and the
consistency of the Ho\v{r}ava model
\ci{Cai:0905,Keha,Chen:0905_2,Char,Li,Soti,Kim,Muko:0905,Gao:0905,
Blas:0906,Koba:0906,Bogd:0907,Wang:0907}. Especially, regarding the
$\lambda=1$ case in which general relativity is expected to be
recovered in IR limit, there have been subtleties in defining
physical modes \ci{Hora:08,Hora,Keha}.

In this paper, I reconsider those problems and show that, in the
Minkowski vacuum background, the extra scalar graviton mode can be
consistently decoupled from the usual tensor graviton modes, by
imposing the (local) Hamiltonian constraint as well as the momentum
constraints. This reduces to the results of Einstein gravity in IR
and achieves the consistency of the model.

To this ends, I start by considering the ADM decomposition of the
metric
\begin{\eq}
ds^2=-N^2 c^2 dt^2+g_{ij}\left(dx^i+N^i dt\right)\left(dx^j+N^j
dt\right)\
\end{\eq}
and the IR-modified Ho\v{r}ava action which reads
\begin{\eq}
S &= & \int dt d^3 x
\sqrt{g}N\left[\frac{2}{\kappa^2}\left(K_{ij}K^{ij}-\lambda
K^2\right)-\frac{\kappa^2}{2\nu^4}C_{ij}C^{ij}+\frac{\kappa^2
\mu}{2\nu^2}\epsilon^{ijk} R^{(3)}_{i\ell} \nabla_{j}R^{(3)\ell}{}_k
-\frac{\kappa^2\mu^2}{8} R^{(3)}_{ij} R^{(3)ij} \right.
\nonumber \\
&&\left. +\frac{\kappa^2 \mu^2}{8(3\lambda-1)}
\left(\frac{4\lambda-1}{4}(R^{(3)})^2-\Lambda_W R^{(3)}+3
\Lambda_W^2\right)+\frac{\kappa^2 \mu^2 \om}{8(3\lambda-1)}
R^{(3)}\right]\ , \label{horava}
\end{\eq}
where
\begin{\eq}
 K_{ij}=\frac{1}{2N}\left(\dot{g}_{ij}-\nabla_i
N_j-\nabla_jN_i\right)\
 \end{\eq}
is the extrinsic curvature (the dot $(\dot{~})$ denotes the
derivative with respect to $t$),
\begin{\eq}
 C^{ij}=\epsilon^{ik\ell}\nabla_k
\left(R^{(3)j}{}_\ell-\frac{1}{4}R^{(3)} \delta^j_\ell\right)\
 \end{\eq}
is the Cotton tensor,  $\kappa,\lambda,\nu,\mu, \La_W$, and $\om$
are constant parameters. The last term, which has been introduced in
\ci{Hora,Keha,Park:0905,Park:0906}, represents a ``soft" breaking of
the ``detailed balance" condition in \ci{Hora} and this modifies the
IR behaviors such that the flat Minkowski vacuum is allowed
\footnote{In \ci{Keha}, $\omega=8 \mu^2(3\lambda-1)/\kappa^2$ has
been considered for the $AdS$ case, but $\om$ may be considered as
an independent parameter, more generally. }.

The action is invariant under the foliation-preserving
diffeomorphism\footnote{This corresponds to a flat slicing of
constant-time surfaces. But more general ({\it space-like}) curved
slicings may be also possible in which the {\it Diff} symmetry with
$\de t =-f (t, {\bf x})$ can be achieved, at least ``formally", with
the corresponding covariant-like actions. (See, for example,
\ci{Germ:0906}). } ({\it Diff})
\begin{\eq}
\label{Diff}
\delta x^i &=&-\zeta^i (t, {\bf x}), ~\delta t=-f(t), \no \\
 \delta
g_{ij}&=&\pa_i\zeta^k g_{jk}+\pa_j \zeta^k g_{ik}+\zeta^k
\pa_k g_{ij}+f \dot g_{ij},\nonumber\\
\delta N_i &=& \pa_i \zeta^j N_j+\zeta^j \pa_j N_i+\dot\zeta^j
g_{ij}+f \dot N_i+\dot f N_i, \no \\
\delta N&=& \zeta^j \pa_j N+f \dot N+\dot f N.
\end{\eq}
Note that this {\it Diff} exists for arbitrary spacetime-dependent
$N,N_i,g_{ij}$. This implies that the equations of motion by varying
$N,N_i,g_{ij}$ are all the ``local'' equations as in the usual
Lorentz invariant Einstein gravity. If we restrict $N$ to be a
function of $t$ only (known as ``projectable" function
\ci{Hora:08,Hora,Soti}), it does not transform to any gauge choice
for which $N$ is a function of space also. So, it seems that there
are two gauge inequivalent classes for Ho\v{r}ava gravity, i.e.,
projectable and non-projectable versions. However, since obtaining
general relativity, including the Newtonian gravity for {\it
generic} gauges, in IR limit could be problematic in the projectable
version and also the choice of $N=N(t)$ can be achieved only for
some limited regions or classes of spacetimes \ci{Soti,Lu,Tang}, we
only consider the non-projetable version in this paper.
%this but this
%does ``not" mean that the equation of motion for $N$, i.e., the
%Hamiltonian constraint equation, is ``not" the local equation ${\cal
%H}^t(t, {\bf x}) \approx 0$ but the global $\int d^3 x {\cal H}^t(t,
%{\bf x}) \approx 0$ which depends on the time $t$ only, as has been
%claimed in the literatures (see for example
%\ci{Hora:08,Hora,Soti}).
\footnote{In (non-projectable) Ho\v{r}ava gravity, the local
Hamiltonian constraint does not form a closed, i.e., first-class
constraint, algebra. However this does not mean that (local)
Hamiltonian constraint can not be imposed consistently but only
means that we have more (secondary) constraints. There have been
some analyses about the additional constraints in the literatures
\ci{Li,Blas:0906} but the full set of the constraints seems to be
still unraveled and deserves fuller investigation.}

Now, in order to study graviton modes, I will consider perturbations
of metric around some appropriate backgrounds, which are solutions
of the full theory (\ref{horava}). But, from the limited knowledge
of the exact (stationary) background solutions\footnote{For an
arbitrary $\La_W$, there is analog of the standard
Schwarzschild-(A)dS solution when considering $\la=1$
\ci{Keha,Park:0905}, but for an arbitrary $\la$ the corresponding
solution is not known yet. In contrast, for (non-stationary)
FRW-type cosmology solution, the vacuum solution for an arbitrary
$\la$ does exist but this can not transform to the stationary form
due to the absence of the full {\it Diff}. }, I consider only the
perturbations around Minkowski vacuum\footnote{The Minkowski vacuum
satisfies trivially the secondary constraint which is generated by
the consistency of the local Hamiltonian constraint
\ci{Li,Blas:0906}.}, which is a solution of the full theory
(\ref{horava}) in the limit of $\La_W \ra 0$,
\begin{\eq}
\label{pert}
 g_{ij}=\de_{ij} +\ep h_{ij}, ~ N=1+ \ep n, ~ N_i = \ep
n_i
\end{\eq}
with a small expansion parameter $\ep$.

From the extrinsic curvatures under the perturbations (\ref{pert}),
\begin{\eq}
K_{ij}&=&\f{\ep}{2} \left(\dot{h}_{ij}- \pa_i n_j- \pa_j n_i \right)
+ {\cal O}
(\ep^2), \no \\
K&=& \f{\ep}{2} \left(\dot{h}-2 \pa_i n^i \right) + {\cal O} (\ep^2)
\end{\eq}
with $ h \equiv \de^{ij} h_{ij}$, the kinetic part $S_K=\int dt d^3
x \sqrt{g}N \frac{2}{\kappa^2}\left(K_{ij}K^{ij}-\lambda K^2\right)$
becomes, at the quadratic order,
\begin{\eq}
\label{S_K}
 S_K=\int dt d^3 x \frac{\ep^2}{2
\kappa^2}\left(\dot{h}_{ij} \dot{h}^{ij}-\lambda \dot{h}^2 -n_i
{\cal H}^i _{(\ep)} \right),
\end{\eq}
where
\begin{\eq}
\label{mom_const} \f{\ep}{\kappa^2} {\cal H}^i _{(\ep)} \equiv -
\f{2 \ep}{\kappa^2} \left[ \pa_t \left( \pa_j h^{ij} -\la \de^{ij}
\pa_j h \right) +(2 \la-1) \pa^i \pa_j n^j -\pa^2 n^i \right]\approx
0
\end{\eq}
are the momentum constraints at the linear order of $\ep$.

On the other hand, the {\it Diff} (\ref{Diff}) reduces to (see
\ci{Keha,Soti} for comparisons)
\begin{\eq}
\label{Diff:linear}
\de x^i &=&- \ep \xi^i (t, {\bf x}), ~\de t =- \ep g(t), \no \\
\de h_{ij} &=&\pa_i \xi_j +\pa_j \xi_i, \no \\
\de n_i &=&\dot{\xi}_i, ~ \de n =\dot{g}.
\end{\eq}
Here, one can choose, by taking time-independent spatial {\it Diff},
$\xi^i =\xi^i ({\bf x })$,
\begin{\eq}
\label{ni:gauge}
 n_i=0
\end{\eq}
but this does not mean the absence of the momentum constraints $\ep
{\cal H}^i_{\ep}\approx 0$ again, as in the $A_0=0$ gauge in the
gauge theory: $A_0$ is the Lagrange multiplier like as $N,~N_i$ and
its variation gives the (local) Gauss' law constraint, but the gauge
choice $A_0=0$ does not mean that there is no {\it local} Gauss' law
constraint \ci{Jack}; indeed, the local Gauss' law is needed in
order to be consistent with the existence of gauge symmetry for $\de
A_i=\pa_i \theta$ independently of the gauge choice $A_0$ and
moreover, the absence of the Gauss' law constraint leads to troubles
in quantization. In this case, one can choose the Ho\v{r}ava's gauge
\ci{Hora:08,Hora} for the perturbed metric $h^{ij}$,
\begin{\eq}
\label{Horava_gauge}
 \pa_j h^{ij} -\la \de^{ij} \pa_j h =0,
\end{\eq}
which is {\it time independent}, according to the momentum
constraints (\ref{mom_const}). Then, the transverse field
\begin{\eq}
H_{ij} \equiv h_{ij} -\la \de_{ij} h,~\pa_i H_{ij}=0
\end{\eq}
may be introduced. This can be further decomposed into its
transverse traceless part $\tilde{H}_{ij}$ and its trace $H=(1-3
\la) h$,
\begin{\eq}
H_{ij} =\tilde{H}_{ij} + \f{1}{2} \left(\de_{ij} -\f{ \pa_i
\pa_j}{\pa^2} \right) H.
\end{\eq}
From these, one obtains
\begin{\eq}
\label{h:decom}
 h_{ij}=\tilde{H}_{ij} + \f{1-\la}{2(1-3 \la)}
\de_{ij} H -\f{1}{2}\f{ \pa_i \pa_j}{\pa^2} H, ~h=\f{H}{1-3 \la}.
\end{\eq}
Then the kinetic part (\ref{S_K}), at the quadratic order of $\ep$,
becomes \ci{Hora:08,Hora}
\begin{\eq}
\label{S_KH}
 S_K=\frac{\ep^2}{2
\kappa^2} \int dt d^3 x \left(\dot{\tilde{H}}_{ij}
\dot{\tilde{H}}^{ij}+\f{1-\la}{2 (1-3 \lambda)} \dot{H}^2 \right).
\end{\eq}

From the intrinsic curvatures \footnote{I follow the conventions of
Wald \ci{Wald}.} under the perturbations (\ref{pert}),
\begin{\eq}
{R^{(3)}}_{ij}&=&\f{\ep}{2} \left( \pa^k \pa_i h_{jk} -\pa^2 h_{ij}
+
\pa^k \pa_j h_{ik} -\pa_i \pa_j h \right) + R^{(NL)}_{ij} %+ {\cal O} (\ep^3)
, \no \\
R^{(3)} &=&\ep \left( \pa_k \pa_i h^{ik} -\pa^2 h \right) +{\cal O}
(\ep^2), \label{R_3}
\end{\eq}
the potential part which is second order in the (spatial)
derivatives in the flat limit $\La_W \ra 0$, $S_{V(2)}=\int dt d^3 x
\sqrt{g}N \frac{\kappa^2 \mu^2 \om}{8(3\lambda-1)} R^{(3)}$ becomes
\begin{\eq}
\label{S_V2}
 S_{V(2)}=-\frac{\ep^2 \kappa^2 \mu^2
\om}{8(3\lambda-1)}\int dt d^3 x \left[ \f{1}{4} h_{ij} \left(
-\pa^2 h^{ij} + 2 \pa^k \pa^i {h^j}_k-2 \pa^i \pa^j h + \de^{ij}
\pa^2 h \right)-n {\cal H}^t_{(\ep)} \right],
\end{\eq}
where
\begin{\eq}
\label{ham_const}
 \ep {\cal H}^t _{(\ep)} \equiv - \ep \pa_k (\pa_i
h^{ik}- \pa^k h) \approx 0
\end{\eq}
is the Hamiltonian constraint\footnote{If one considers the last
term in (\ref{S_V2}), $-n {\cal H}^t_{(\ep)}$, as $ h^{ik} \pa_k
\pa_i n -h \pa^2 n$, by taking the integration by parts, and solves
the equations of motions for $h_{ij}$ first, a lot of troubles
occur. Actually, this is the source of the troubles in
\ci{Blas:0906}. However, by considering the time-independent
temporal {\it Diff}, i.e., $g=constant$, one can always choose the
gauge, called synchronous or (gravitational) Weyl gauge together
with (\ref{ni:gauge}), $n=0$ also and the troublesome terms
disappear; this implies that the strong coupling problem which has
been observed in \ci{Blas:0906} would be a gauge artifact. This
result agrees with the perturbations in FRW background
\ci{Gao:0905}.} at the linear order of $\ep$. Here, I have used
\begin{\eq}
\sqrt{g}  R^{(3)}&=&\de^{ij} R_{ij}^{(NL)} +\ep h_{ij} \left(-R^{
ij (L)} +\f{1}{2} \de^{ij} R^{(L)} \right) + {\cal O} (\ep^3) \no \\
&=&\f{\ep}{2} h_{ij} \left(-R^{ij (L) } +\f{1}{2} \de^{ij} R^{(L)}
\right) + {\cal O} (\ep^3),
\end{\eq}
where $R^{(L)}_{ ij}, R^{(NL)}_{ ij}$ denote the linear, non-linear
perturbations of $R^{(3)}_{ij}$ in (\ref{R_3}), respectively. The
action (\ref{S_V2}), when combined with the Ho\v{r}ava's gauge
(\ref{Horava_gauge}), reduces to
\begin{\eq}
\label{S_V2b}
 S_{V(2)}=\frac{\ep^2 \kappa^2 \mu^2
\om}{8(3\lambda-1)}\int dt d^3 x \left[ \f{1}{4} h_{ij} \pa^2 h^{ij}
+ \f{(2 \la(1-\la)-1)}{4} h \pa^2 h +n {\cal H}^t_{(\ep)} \right].
\end{\eq}

On the other hand, the Hamiltonian constraint (\ref{ham_const}),
when combined with the gauge fixing condition (\ref{Horava_gauge}),
reduces to
\begin{\eq}
\label{ham_const_red}
 (\la -1) \pa^2 h \approx 0.
\end{\eq}
For $\la \neq 1$, this leads to
\begin{\eq}
\pa^2 h \approx 0
\end{\eq}
but, for $\la =1$, (\ref{ham_const_red}) is automatically satisfied.
This is basically due to the fact that the momentum and Hamiltonian
constraints (\ref{mom_const}) and (\ref{ham_const}), ``degenerate"
for the Ho\v{r}ava's gauge (\ref{Horava_gauge}) with $\la=1$, at the
linear order of $\ep$. In other words, for $\la=1$ the gauge fixing
condition (\ref{Horava_gauge}) is consistent only if the (local)
Hamiltonian constraint (\ref{ham_const}) is considered; with the
Hamiltonian constraint (\ref{ham_const}), the gauge condition
(\ref{Horava_gauge}) can be consistent for arbitrary values of
$\la$, including $\la=1$. Actually, this has been a source of some
confusions and troubles in the literatures. For example, in the
projectable version \ci{Hora:08,Hora,Soti}, the global Hamiltonian
constraint $\int d^3 x {\cal H}^t \approx 0$ has been considered
together with the momentum constraint (\ref{mom_const}), but in this
case, there has been a problem in defining the gauge condition
(\ref{Horava_gauge}) for $\la=1$: Applying $\pa_i$ to
(\ref{Horava_gauge}), the left hand side equals to the linearized
Ricci scalar $R^{(3)}$ in (\ref{R_3}) which can not set to zero by
gauge transformation with the global Hamiltonian constraint $\int
d^3 x R^{(3)} \approx 0$, generally \footnote{The global constraint
$\int d^3 x {\cal H}^t \approx 0$ in \ci{Hora:08,Hora,Soti} produces
the equations for spatial infinity, in the absence of the inner
boundary, due to the total derivative form of (\ref{ham_const}) at
the linear order of $\ep$. But this can be negligible for the fields
$h_{ij}$ which decay fast enough at infinity. }.

From the mode decomposition, the second-order spatial derivative
action (\ref{S_V2b}) becomes
\begin{\eq}
 S_{V(2)}=\frac{\ep^2 \kappa^2 \mu^2
\om}{8(3\lambda-1)}\int dt d^3 x \left[ \f{1}{4} \tilde{H}_{ij}
\pa^2 \tilde{H}^{ij} -\f{(1-\la) (1+3 \la)}{8 (1-3 \la)^2} H \pa^2 H
+n {\cal H}^t_{(\ep)} \right]
\end{\eq}
with
\begin{\eq}
\label{ham_const_H}
 {\cal H}^t_{(\ep)}=\f{1-\la}{(1-3 \la)} \pa^2 H \approx 0.
\end{\eq}
Then the second-order derivative action becomes altogether
\begin{\eq}
 S_{(2)}&=& \ep^2\int dt d^3 x \left[\frac{1}{2 \kappa^2} \dot{\tilde{H}}_{ij}
\dot{\tilde{H}}^{ij}+ \frac{ \kappa^2 \mu^2 \om }{32(3\lambda-1)}
\tilde{H}_{ij} \pa^2 \tilde{H}^{ij}
\right. \no \\
&& \left. +\frac{1-\la}{4 (1-3 \la) \kappa^2} \dot{H}^2 +\f{
\kappa^2 \mu^2 \om (1-\la) (1+3 \la)}{64 (1-3 \la)^3} H \pa^2 H +
\frac{ \kappa^2 \mu^2 \om}{8(3\lambda-1)}n {\cal H}^t_{(\ep)}
\right].
\end{\eq}
The first two terms represent the usual transverse traceless
graviton modes $\tilde{H}_{ij}$ with the speed of gravitational
interaction
\begin{\eq}
c_g=\sqrt{ \f{\kappa^4 \mu^2 \om}{16 (3 \la -1)}},
\end{\eq}
which agrees with the speed of light $c$ in the IR and $\La_W \ra 0$
limits of the action (\ref{horava}) \ci{Keha,Park:0906}:
\begin{\eq}
S_{\la \rm EH} =  \f{c^4}{16 \pi G } \int dt d^3 x
\sqrt{g}N\left[\frac{1}{c^2}\left(K_{ij}K^{ij}-\lambda
K^2\right)+R^{(3)} \right]. \label{lEH}
\end{\eq}
Here it is important to note that the propagation can exist due to
the IR modification term with an arbitrary coefficient $\om$, which
has been overlooked in \ci{Hora:08,Hora} but corrected in \ci{Keha}.
The next two terms seem to imply another scalar mode $H$ but this
depends on the values of $\la$: For $\la \neq 1$, this mode is
physical but non-propagating in the physical subspace of the
Hamiltonian constraint (\ref{ham_const_H}), giving $ \pa^2 H \approx
0$. On the other hand, for $\la=1$, where the Hamiltonian constraint
is trivially satisfied due to the degeneracy with the momentum
constraints, the mode $H$ is completely disappeared in the action
and this agrees with the usual Einstein gravity. Actually, this can
be more easily understood in the decomposition (\ref{h:decom}) in
which the second term is absent for $\la=1$ and then the remaining
term of $H$ can be gauged away due to the symmetry
(\ref{Diff:linear}). \footnote{In some literatures
\ci{Hora:08,Hora,Cai:0905,Soti}, it was claimed that the equation of
motion of the scalar mode for $\lambda=1$ reduces to $\ddot{H}=0$,
giving the linearly expanding solution $H(t,{\bf x})=H_0({\bf x}) +
t H_1 ({\bf x})$, but this undesirable mode can be eliminated by the
extra gauge invariance for $\lambda=1$ such that the usual general
relativity is recovered in IR. However, this argument is quite
subtle since the correct equation of motion is
$(\lambda-1)\ddot{H}=0$, which is trivially satisfied for arbitrary
solution of $H(t,{\bf x})$ which can go beyond the extra gauge
transformation. } This provides a consistency of the Ho\v{r}ava
gravity in the IR limit.

The UV behaviors are governed by the higher derivative terms in
(\ref{horava}) and the quadratic part of the perturbed action is
\begin{\eq}
\label{S_UV}
 S_{(UV)}&=& \f{\ep^2}{4} \int dt d^3 x \left[ -\bar{a}  \tilde{H}_{ij} \pa^6
\tilde{H}^{ij} + \bar{b}  \ep^{ijk} \tilde{H}_{il} \pa^4 \pa_j
{\tilde{H}^l}_{k} +\bar{c} \tilde{H}_{ij} \pa^4 \tilde{H}^{ij} \right.\no \\
&&~~~~~~~~~~~~~\left.+\f{(\la-1)^2}{(1-3 \la)^2} \left( \f{3
\bar{c}}{2} + 4 \bar{d} \right)H \pa^4 H \right],
\end{\eq}
where
\begin{\eq}
\label{detailed}
 \bar{a}=-\f{\kappa^2}{2 \nu^4}, ~ \bar{b}=\f{\kappa^2 \mu}{2
\nu^2},~\bar{c}=-\f{\kappa^2 \mu^2}{8},~ \bar{d}=\f{\kappa^2 \mu^2
(4 \la-1)}{32 (3 \la-1)}
\end{\eq}
are the coefficients of $C_{ij} C^{ij},~\epsilon^{ijk}
R^{(3)}_{i\ell} \nabla_{j}R^{(3)\ell}{}_k ,~ R^{(3)}_{ij}
R^{(3)ij}$, and $R^{(3)} R^{(3)}$, respectively. The first three
terms provide the modified dispersion relation $\om^2 \sim k^6+
\cdots$ for the transverse traceless modes. Here, the (UV) detailed
balance with the particular values of the coefficients
(\ref{detailed}) do not have any role. The last term contains higher
spatial derivatives of the scalar mode $H$ but this does not appear
in the physical subspace of either $\la \neq 1$, giving $\pa^2 H
\approx 0$, or $\la=1$, again. Here, the non-existence of sixth
derivative terms for the scalar mode is the results of the detailed
balance in sixth order,
\begin{\eq}
C_{ij} C^{ij} = \al \nabla_{i}R^{(3)}_{jk} \nabla^{i}{R^{(3)}}^{jk}
+\be \nabla_{i}R^{(3)}_{jk}\nabla^{j} {R^{(3)}}^{ik}
+\ga\nabla_{i}R^{(3)}\nabla^{i}R^{(3)}
\end{\eq}
with $\al=1,\be=-1,\ga=-1/8$. On the other hand, for arbitrary
values of $\al,\be,\ga$ one obtains
\begin{\eq}
C_{ij} C^{ij} = -\f{\al \ep^2 }{4} \tilde{H}_{ij} \pa^6
\tilde{H}^{ij} -\f{\ep^2 (\la-1)^2}{4 (1-3 \la)^2} \left( \f{3
\al}{2} +\be +4 \ga \right) H \pa^6 H
\end{\eq}
and there are sixth derivative terms for the scalar mode $H$. But,
even in this case, these terms do not produce the propagation in the
physical subspace, for arbitrary values of $\la$.

In conclusion, I have reconsidered the problem of the extra scalar
graviton mode in Ho\v{r}ava gravity. I showed that, in the Minkowski
vacuum background, the scalar mode excitation can be consistently
decoupled from the usual tensor graviton modes in UV as well as in
IR, by imposing the (local) Hamiltonian constraint as well as the
momentum constraints, regardless of $\la=1$ or not. This provides a
consistency of the IR modified Ho\v{r}ava gravity for the quadratic
perturbations in the Minkowski vacuum background. It would be
interesting to study the role of the local Hamiltonian constraint in
the scalar mode decoupling with the more general backgrounds with
matters and higher order perturbations which have been also debating
issues \ci{Char,Muko:0905,Gao:0905,Blas:0906}.

{\it Note Added}: After the appearance of this paper, there have
been several analyses on the number of physical modes of Ho\v{r}ava
gravity through the constraints algebra. First at the linear order
for cosmological perturbations, it has been found \ci{Gong} that
there is no additional scalar perturbation mode, in agreement with
\ci{Gao:0905} and the present paper. Later at the fully non-linear
orders for the IR limit of Ho\v{r}ava gravity (\ref{lEH}), it has
been also found that the number of physical degrees of freedom is
the same as GR, which implies that there is no non-perturbative
generation of scalar graviton as well, in the IR limit. \ci{Bell}

\section*{Acknowledgments}

I would like to thank Li-Ming Cao, Shinji Mukohyama, Yun Soo Myung
for helpful discussions. This work was supported by the Korea
Research Foundation Grant funded by Korea Government(MOEHRD)
(KRF-2007-359-C00011).

%%%%%%%%%% References %%%%%%%%%%%%%%%%%%%%%%%%%
\newcommand{\J}[4]{#1 {\bf #2} #3 (#4)}
\newcommand{\andJ}[3]{{\bf #1} (#2) #3}
\newcommand{\AP}{Ann. Phys. (N.Y.)}
\newcommand{\MPL}{Mod. Phys. Lett.}
\newcommand{\NP}{Nucl. Phys.}
\newcommand{\PL}{Phys. Lett.}
\newcommand{\PR}{Phys. Rev. D}
\newcommand{\PRL}{Phys. Rev. Lett.}
\newcommand{\PTP}{Prog. Theor. Phys.}
\newcommand{\hep}[1]{ hep-th/{#1}}
\newcommand{\hepp}[1]{ hep-ph/{#1}}
\newcommand{\hepg}[1]{ gr-qc/{#1}}
\newcommand{\bi}{ \bibitem}
%%%%%%%%%%%%%%%%%%%%%%%%%%%%%%%%%%%%%%%%%%%%%%%

\end{document}